\documentclass[english,twocolumn,showpacs,preprintnumbers,amsmath,amssymb]{revtex4}
\usepackage[T1]{fontenc}
\usepackage[latin9]{inputenc}
\usepackage{amsmath}
\usepackage{graphicx}
\usepackage{amssymb}

\makeatletter
\@ifundefined{textcolor}{}
{%
 \definecolor{BLACK}{gray}{0}
 \definecolor{WHITE}{gray}{1}
 \definecolor{RED}{rgb}{1,0,0}
 \definecolor{GREEN}{rgb}{0,1,0}
 \definecolor{BLUE}{rgb}{0,0,1}
 \definecolor{CYAN}{cmyk}{1,0,0,0}
 \definecolor{MAGENTA}{cmyk}{0,1,0,0}
 \definecolor{YELLOW}{cmyk}{0,0,1,0}
 }



\usepackage{babel}

\begin{document}

\title{Electron shell and the $\alpha$-decay}

\author{$^1$Sergey Yu. Igashov and $^2$Yury M. Tchuvil'sky}

\affiliation{$^1$All-Russia Research Institute of Automatics, 127055, Moscow, Russia\\
$^2$Skobeltsyn Institute of Nuclear Physics, Lomonosov Moscow State University, 119991,
Moscow, Russia}

\date{}
\begin{abstract}
The ratio of the $\alpha$-decay widths of a bare nucleus and the related atom is
calculated. Both the change of the form and thus the penetrability of the potential
barrier and the effect of reflection in the classically-allowed region appearing due to
the electron shell are taken into account in the calculations of this ratio. The
contribution of each of these two effects is of one and the same order of magnitude. For
long-lived radioactive samples the values of the total effect turn out to be somewhat
below 1 percent.
\end{abstract}



\maketitle

\section {Introduction}

Transmutation of long-lived radioactive isotopes is one of the hottest problem of the
modern nuclear physics. Recent attempts to produce an affect upon the decay rate of such
isotopes by physical methods which do not deal with any beam of neutrons, gamma-quanta
or charged particles demonstrate increasing interest and popularity of this problem. By
now a list (not so wide) of examples demonstrating the possibilities to vary the rate of
$\beta$-transitions by deep ionization of atoms is known \citep{Tak, Jung}. Also it is
shown that the half-life times of some low-energy $\gamma$-transitions turn out to be
much smaller when the energy of the nuclear transition is close to the energy of any
allowed atomic transition of the same multipolarity \citep{Karp,Tk}. The $\alpha$-decay
looks as a process which is the most insensitive to the external influence of the
discussed type due to the typical value of the energy of the potential barrier which is
in the neighborhood of 20 MeV. That is why this process attracts the smaller interest
than other types of decay. Nevertheless in the last decade some theoretical papers
devoted to this problem were published \citep{Zinn, Bar}. Moreover an interesting
supporting evidence is obtained in the experiment presented in Ref. \citep{Mikh}. The
difference between the decay widths of $^{212}$Po in Pb and Ni matrixes is proved there
with 95 per cent reliability (see the discussion bellow). Thus, in our opinion, it is
necessary to study accurately the effects modifying the $\alpha$-decay widths.

In the present paper the $\alpha$-decay widths of a bare nucleus and the respective atom
are compared. The electron cloud changes, first, the kinetic energy of the emitted
$\alpha$-particle and, second, the decay rate. The effect of the kinetic energy
difference in cases of the decays of the bare nucleus and the atom is well-known from
the middle of the last century. As usual nuclear data tables contain the values of
energy yield of the $\alpha$-decay of atoms. Therefore the effect can not be neglected
and should be accurately considered: the formal inclusion of such energy correction
leads to increasing of the $\alpha$-decay width by the factor lying between 1.22 (for
$^{212}$Po) and 2.60 (for $^{147}$Sm) \citep{Erma}. This circumstance is always taken
into account. The problem how the decay rate is changed by the electron cloud has not
solved accurately up to now.

A closely related problem is examined in Ref. \citep{Zinn}. To some extent paper
\citep{Zinn} is a reaction of scientific community to mass media publications in which
the $\alpha$-decay enhancement is anticipated in the conditions, similar to that
discussed by the proponents of the so-called "cold fusion". The paper provides a
qualitative argument against these anticipations. It is shown that the upper limit of
the ratio of the $\alpha$-decay half-lives of a nucleus screened by electrons of a
surrounding dense medium and the respective unscreened nucleus in actual cases is not
larger than 1.02. At the same time the author does not decide between two results
obtained for two versions of the screening ($\sim1.013$ and 1.000 in $^{226}$Ra) i. e.
between the modest and zero effects. Some other details of the investigations which is
carried out in Ref. \citep{Zinn} are discussed bellow. So in our opinion the outlined
above problem has been unresolved till now.

\section{Statement of the problem and formalism}

It is convenient to begin consideration of the problem with a simple quasi-classical
approximation. The potential of the interaction of the $\alpha$-particle with the atomic
shells is monotonically increasing negative function $V_{sh}(r)$, very slowly varying in
the potential barrier region. Neglecting the $r$-dependence of the potential in this
region (assuming $V_{sh}(r)=-V_0$), one obtains the only effect of the shift (equal to
$V_0$) of the decay energy. As it is pointed out in Ref. \citep{Zinn} the $\alpha$-decay
width remains the same. A small increase of the $V_{sh}(r)$ results in increasing of the
barrier height and consequently in the decreasing of the barrier penetrability which
takes the form:
\begin{equation}
P_{at}\!=\!exp\left[\!-\frac{1}{\hbar}\!\int\limits_{r_{int}}^{r_{ext}}
\!\!\!{\sqrt{V_{bn}(r)\!+\!V_{sh}(r)\!-\!E_{bn}\!+\!V_0}}\,dr\right],
\label{pen}\end{equation}
where $E_{bn}$ is the resonance energy in the bare nucleus case,
$V_{bn}(r)=V_{str}(r)+V_{cf}+V_{Coul}(r)$ is the sum of the strong, the centrifugal, and
the Coulomb potentials of the  interaction of the $\alpha$-particle and the residual
nucleus.

At the same time there is another effect originated by the atomic shell. It appears due
to the fact that the asymptotic behavior of the resonance (Gamow) wave function in the
case of bare nucleus differs from that of the respective atom. Indeed, the asymptotic
behavior of a resonant function looks as follows:
\begin{equation}
\chi_l^{res}(r)=D\bigr[G_l(\eta,kr)+{\rm i}F_l(\eta,kr)\bigl],\ \ r\to \infty,
\label{as}\end{equation}
where $G_l(\eta,\rho)$ and $F_l(\eta,\rho)$ are the Coulomb wave functions
\citep{Abramowitz},
\begin{equation}
\eta=(e_0^2/\hbar c) Z_1Z_2\sqrt{\mu c^2/(2E)} \label{et}\end{equation}
is the Coulomb parameter, $e_0>0$ -- the elementary charge, $E$ -- the decay energy
yield, $k$ -- the respective wave number, and $\mu$ -- the reduced mass, $Z_2=2$ -- the
$\alpha$-particle charge. The coefficient
\begin{equation}
D=\sqrt{\mu\Gamma}/\hbar, \label{pen1}
\end{equation}
determines the decay width $\Gamma$. In the case of bare nucleus the parameters
$D_{bn}$, $\eta_{bn} $, and $k_{bn}$ are related to the energy $E_{bn}$ and the charge
of the daughter nucleus $Z_1=Z-2$ whereas the corresponding parameters $\eta_{at} $ and
$k_{at}$ (the case of neutral atom) are related to the energy $E_{at}=E_{bn}-V_0$ and
the charge $Z_1=-2$ of the residual system -- the nucleus and the electron cloud. The
origin of the discussed effect is the reflection of the $\alpha$-particle wave on the
varying potential. An obvious way to account for this effect is to solve the
Schr\"odinger equation directly. It may be found grater or smaller than the effect of
the barrier form variation considered above. The main goal of the present paper is to
calculate the ratio of the decay width, accounting for both these effects.

A physical model applied here for the description of the $\alpha$-decay of the atomic
system includes the following ingredients.

First, the distribution of the electronic cloud is assumed to be undistorted during the
$\alpha$-particle emission. This assumption is based on the facts that the orbits of the
strongly bound electrons are slightly changed due to the variation of the nuclear charge
$Z \to Z-2$ and the weakly bound electrons are too slow. Indeed, the electron velocity
$v_e\approx v_{\alpha}$ ($v_{\alpha}$ is a velocity typical for the $\alpha$-particle)
corresponds to the electron energy $E_e\simeq 500$ eV. Thus a high speed and a small
charge of the emitted particle enables one to represent the potential energy term of the
Hamiltonian in the folding-form interaction of the $\alpha$-particle with negative ion
consisting of the daughter nucleus and the atomic shell of the mother one.

Second, the electron distribution is taken in the Thomas-Fermi form \citep{TF}.

Third, the electron charge, associated with sphere of the radius $R_0$ which is somewhat
greater than that of the range of strong interaction $R_{str} \sim [1.2 A^{1/3} +
R_{\alpha} + 2] \quad {\rm fm}$, is assumed to be small enough to neglect the variation
of the potential originated by this charge.

At last only the ratio of the $\alpha$-decay widths but not the absolute values of the
widths for both an atom and a bare nucleus is the object of the calculation. Evidently
the absolute values are related to many-nucleon properties of a nucleus and thus are
beyond the scope of the present paper.

The properties of the Gamov solutions to the two-body Schr\"odinger equation related to
the presented model are the following. The solution is characterized by the asymptotic
form \eqref{as} with the parameters $D_{at}$, $\eta_{at} $, and $k_{at}$ at very large
(several times grater than the atomic radius) distances. At smaller distances it can be
obtained only numerically. In the region $R_{str}<r<R_0$ the equation turns out to be
the Coulomb one with the parameters $\eta_{bn} $ and $k_{bn}$ related to the bare
nucleus. Consequently in this domain the resonance solution becomes:
\begin{equation}
\chi_l^{res}(r)=D_{at}\bigr[AG_l(\eta_{bn},k_{bn}r)+BF_l(\eta_{bn},k_{bn}r)\bigl].
\label{fint}\end{equation}
In this region the coefficients $D_{at}$, $A$, and $B$ differ the solution from the
one, corresponds to the Gamow state of bare nucleus, where $D=D_{bn}$, $A=1$,
and $B={\rm i}$ (see Eq. \eqref{as}).

In the both cases the amplitude of each wave function is determined by the matching of
the function of the type \eqref{fint} with one and the same internal wave function. As
it is shown in the Ref. \citep{Kadm} the term containing the regular Coulomb function
plays a negligible role in the decay width calculations of a narrow resonance. Indeed,
the ratio of the regular and the irregular Coulomb wave functions in the sub-barrier
region can be estimated as follows:
$$F_l(\eta_{bn}, k_{bn}r)/G_l(\eta_{bn}, k_{bn}r)\cong P(r) = $$
\begin{equation}
exp \left[-\frac{1}{\hbar}\int\limits_r^{r_{ext}}
{\sqrt{E-V(\rho)}}\,d\rho \right],\ \ r\ll r_{ext}.
\label{fg}\end{equation}
Remind that for the $\alpha$-emitters with the half-lives $\tau> 1$ y the value of the
total penetrability of the barrier (which is expressed by Eq. \eqref{fg} with the lower
limit $r=r_{int}$) $P < 10^{-27}$. Furthermore, the region beyond the radius $R_{str}$
contributes predominantly to the value of the integral in Eq. \eqref{fg}. Thus, the
second term in Eq. \eqref{fint} can be neglected with a high precision. Therefore the
squared ratio of the amplitudes of the wave functions and consequently the ratio of the
widths become
\begin{equation}
\Gamma_{bn}/\Gamma_{at}=|D_{bn}/D_{at}|^2=|A|^2
\label{subb}
\end{equation}

This coefficient should be calculated numerically. The calculation procedure is
complicated due to the need to solve numerically Schr\"odinger equation in extremely
wide domain of values of the radial variable $r$. The results of the performed
calculations demonstrate that the asymptotic form \eqref{as} becomes valid at the
distance of the order of $10^6$ fm whereas a typical value of the wave number of the
$\alpha$-particle is about $k\simeq1$ fm$^{-1}$. At the same time a high numerical
precision should be provided both at the large distance and in the neighborhood of the
nucleus. In fact one deals with two very different scales of distances in one and the
same calculation procedure.

The optimal way is as follows. As it is noted above one has no need to solve the
Schr\"odinger equation in the $r\leq R_{str}$ region for the calculation of the ratio of
the decay widths of an atom and the respective bare nucleus -- this ratio is determined
beyond the region. Therefore it is convenient to consider the reduced equation:
\begin{equation}
\biggl(-\frac{\hbar^2}{2\mu}\frac{d^2}{dr^2}+V_{Coul}(r)+V_{sh}(r)+V_{cf}(r)-E\biggr)
\chi_l(r)=0.
\label{schr}\end{equation}

The Thomas-Fermi model was utilized for description of the electron cloud potential:
\begin{equation}
V_{sh}(r)=2\varphi_{sh}(r)e_0, \label{9}\end{equation}
\begin{equation}
\varphi_{sh}(r)={\frac{-Ze_0}{r}}(1-\xi(x)), \label{10}
\end{equation}
$x={\displaystyle rZ^{1/2} m_ee_0^2b^{-1}\hbar^{-2}}$, $b={\displaystyle
(3\pi/4)^{2/3}/2}$ ($m_e$ -- electron mass). In this case $V_0=2m_eZ^{4/3}e_0^4\kappa
b^{-1}\hbar^{-2}$ and $\kappa\approx1.59$ is the constant in the asymptotic formula
$\xi(x)\simeq1-\kappa x$, as $x\rightarrow 0$. We introduce the distance $R_0$ in such a
way that in the domain $r<R_0$ the $r$-dependence of $V_{sh}$ can be neglected:
$V_{sh}(r)\cong -V_0$. Thus in the domain $R_{str}<r<R_0$ equation \eqref{schr} takes
the form
\begin{equation}
{\frac{d^2\chi_l(r)}{dr^2}}+\biggl({\frac{2\mu E_{bn}}{\hbar^2}}-
{\frac{4\mu}{\hbar^2}}{\frac{(Z-2)e_0^2}{r}} - {\frac{l(l+1)}{r^2}}\biggr)\chi_l(r)=0.
\label{11}\end{equation} This equation coincides with that in the case of bare nucleus
but with shifted energy $E_{bn}$.

The Coulomb wave functions $F_l(\eta_{bn},k_{bn}r)$ and $G_l(\eta_{bn},k_{bn}r)$ are the
solutions to Eq. \eqref{11}. Thus the two linearly independent solutions to Eq.
\eqref{schr} can be chosen in the following way
\begin{equation}
\chi_l^{(F)}(r)=F_l(\eta_{bn},k_{bn}r),\ \ R_{str}<r<R_0, \label{12}\end{equation}
\begin{equation}
\chi_l^{(G)}(r)=G_l(\eta_{bn},k_{bn}r),\ \ R_{str}<r<R_0. \label{13}\end{equation}

At large distances, far from the electron shells the sum of potentials of electron
shells and nucleus looks like the potential of negative ion with the charge $Z_1=-2$. In
this region the solutions to Eq. \eqref{schr} have the form of linear combinations of
functions $F_l(\eta_{at},k_{at}r)$ and $G_l(\eta_{at},k_{at}r)$:
\begin{equation}
\chi_l^{(F)}(r)=\alpha F_l(\eta_{at},k_{at}r)+\beta G_l(\eta_{at},k_{at}r),
\label{14}\end{equation}
\begin{equation}
\chi_l^{(G)}(r)=\gamma F_l(\eta_{at},k_{at}r)+\delta G_l(\eta_{at},k_{at}r).
\label{15}
\end{equation}
The reversed relationships
\begin{equation}
F_l(\eta_{at},k_{at}r)=[\delta\chi_l^{(F)}(r)-\beta\chi_l^{(G)}(r)]/\Delta, \label{16}
\end{equation}
\begin{equation}
G_l(\eta_{at},k_{at}r)=[-\gamma\chi_l^{(F)}(r)+\alpha\chi_l^{(G)}(r)]/\Delta,
\label{17}
\end{equation}
\begin{equation}
\Delta=\alpha\delta-\beta\gamma \label{18}
\end{equation}
follow directly from Eqs. \eqref{14} and \eqref{15}. The value of $\Delta$ can be found
without explicit values of $\alpha$, $\beta$, $\delta$, $\gamma$. Indeed, the
relationship
\begin{equation}
W_r\bigl(\chi_l^{(F)},\chi_l^{(G)}\bigr)=W_r\bigl(F_l(\eta_{at},k_{at}r),
G_l(\eta_{at},k_{at}r)\bigr)\Delta \label{19}\end{equation} follows directly from
\eqref{14}, \eqref{15}, and \eqref{18}.
The lower index $r$ means that the derivatives
with respect to variable $r$ appear in the Wronskian $W_r$. As it is known
\citep{Abramowitz}
\begin{equation}
W_{\rho}\bigl(F_l(\eta,\rho),G_l(\eta,\rho)\bigr)=1, \label{20}
\end{equation}
so from \eqref{12}, \eqref{13}, and \eqref{19} it follows that
\begin{equation}
\Delta=k_{bn}/k_{at}. \label{21}\end{equation} In the case of $\alpha$-decay of a
neutral atom the resonant wave function has the asymptotic form
$D_{at}\bigl[G_l(\eta_{at},k_{at}r)+\rm iF_l(\eta_{at},k_{at}r)\bigr]$ at large
distances far from the electron shells. Now, with the aid of \eqref{16}, \eqref{17} it
is clear that the considered resonance wave function looks as follows
\begin{equation}
\chi^{res}_l=D_{at}\biggl[{\frac{\alpha-{\rm i}\beta}{\Delta}}\chi_l^{(G)}+ {\rm
i}{\frac{\delta+{\rm i}\gamma}{\Delta}}\chi_l^{(F)}\biggr]. \label{22}\end{equation} The
function \eqref{22} in the domain $R_{str}<r<R_0$ has the form (see \eqref{12},
\eqref{13}):
\begin{equation}
D_{at}\biggl[{\frac{\alpha-{\rm i}\beta}{\Delta}}G_l(\eta_{bn},k_{bn}r)+
{\rm i}{\frac{\delta+{\rm i}\gamma}{\Delta}}F_l(\eta_{bn},k_{bn}r)\biggr].
\label{23}\end{equation}

Thus, we come to the following conclusion. In the case of the neutral atom the internal
wave function must be matched with the function \eqref{23}, while in the case of the
bare nucleus with the function $D_{at}\bigl[G_l(\eta_{bn},k_{bn}r)+{\rm
i}F_l(\eta_{bn},k_{bn}r)\bigr]$. The matching procedure can be realized at any point of
the domain $R_{str}<r<R_0$.

In the case of neutral atom the multiplier of $G_l(\eta_{bn},k_{bn}r)$ in Eq.
\eqref{23}, which is symbolized as $A$ in Eqs. \eqref{fint} and \eqref{subb}, changes
the width of the resonant state. Obviously there is no need to solve the resonant
problem as a whole to determine the coefficient $A$. It is sufficient to solve
(numerically) the Schr\"odinger equation \eqref{schr} in the domain $r>R_{str}$ with the
boundary condition \eqref{12}. Then the coefficients $\alpha$ and $\beta$ can be deduced
from the Wronskians of the solution and the functions $G_l(\eta_{at},k_{at}r)$ and
$F_l(\eta_{at},k_{at}r)$. The explicit expression
\begin{equation}
\Gamma_{bn}/\Gamma_{at}=\vert(\alpha-{\rm i}\beta)/\Delta\vert^2=
(\alpha^2+\beta^2)/\Delta^2 \label{24}\end{equation}
for the suppression factor -- the
ratio of the decay width of the bare nucleus and the neutral atom -- follows directly
from \eqref{pen1} and \eqref{fint}. It should be pointed out that we arrived at a
surprising conclusion. Indeed, one needs to find the solution with boundary condition
\eqref{12} (regular function $F_l(\eta_{at},k_{at}r)$) to calculate the coefficient in
the term containing the irregular function $G_l(\eta_{at},k_{at}r)$.

\section {Results and discussion}

The influence of the electron surrounding onto the $\alpha$-decay width was investigated
for some isotopes in the framework of the rigorous approach described above.
\begin{table}[h]
\caption{The values of the coefficient of suppression of the $\alpha$-decay widths by
the electron shell.} \label{Gam}
\begin{center}
\begin{tabular}{c c c c}
\hline \hline Isotope & $Q_{at}$ & $\Gamma_{bn}/\Gamma_{at}$& $P_{bn}/P_{at}$\\
\hline $^{226}$Ra & 4.7806 & 1.0021 & 1.006 \\
 $^{232}$Th & 4.0288 & 1.0052 & 1.010 \\
 $^{144}$Nd & 1.9251 & 1.0079 & 1.014 \\
 $^{148}$Sm & 1.9858 & 1.0085 & 1.015 \\
\hline\hline
\end{tabular}
\end{center}
\end{table}
The Runge-Kutta and Stoermer methods were applied to solve numerically the radial
Schr\"odinger equation \eqref{schr}. The results obtained by means of these methods
coincide with high accuracy. The factors of the $\alpha$-decay suppression caused by the
electron shells are presented in Table I. The third column of the table demonstrates the
results of precise calculations and the last one the results of quasiclassical
calculations of the barrier penetrabilities. The results of exact calculations, based on
Eq. \eqref{24} differ from that of quasiclassical approach. The reason of this
discrepancy is the following. The ratio $P_{bn}/P_{at}$ is the exponential function of
difference of two integrals (see \eqref{pen}) over the subbarrier regions. It turns out
that the vicinity of the corresponding external turning points (upper limits of the
integrals) contributes predominantly to the difference of the quasiclassical integrals.
But the vicinity of a turning point is just the region, where the quasiclassical
approximation looses accuracy.

In a qualitative sense the results presented in the table are not in contradiction with
the experimental one \cite{Mikh} where the following difference between the half-life
times of $^{212}$Po decay in Pb and Ni matrixes is obtained: $T_{1/2}({\rm
Pb})-T_{1/2}({\rm Ni})=(-0.66 \pm 0.25)\ {\rm ns}.$ This difference, related to the
half-life time of $^{212}$Po isotope $T_{1/2}=0.3$ $\mu{\rm s}$ is about 0.2 per cent.
It should be stressed that the settings of the problems in the present work and in Ref.
\cite{Mikh} are different. The effect of the change of solid state surrounding of the
emitter but not the effect of the ionization is studied in the experiment. So, a direct
comparison is impossible.

As it is noted above the upper limit of an $\alpha$-decay half-life time variation in a
dense (solid or liquid) medium is estimated theoretically in Ref. \cite{Zinn}. The
effect is assumed to be originated by the change of the quasiclassical penetrability of
the barrier by the potential of the interaction of the $\alpha$-particle with electrons
of the medium. The initial barrier is chosen in the simple form of "cut Coulomb
potential". The change of the barrier is described by the exponential screening factor
$exp(-r/R)$. Two parameters $R=3.7\cdot10^3$ fm (for so-called "Thomas-Fermi screening")
and $R=4\cdot10^5$ fm (for "Debye screening") are considered. The values
$P_{bn}/P_{at}=1.013$ and $1.000$ for $^{226}$Ra are obtained in these cases
respectively. The larger value of the upper limit is qualitatively confirmed in our
calculations.

So the ratio of the $\alpha$-decay widths of a bare nucleus and the related atom is
calculated in accurate way. The mathematics providing a possibility to take into account
the effect of reflection of the $\alpha$-particle wave at the electron shell in the
classically-allowed region is built. The change of the penetrability of the potential
barrier is taken into account simultaneously.  For long-lived radioactive samples the
values of the effect turn out to be somewhat below 1 percent. They decrease with
increasing of the decay energy.

The authors are thankful to V. G. Kalinnikov for motivating discussion.

\end{document}